\documentclass[useAMS,usenatbib]{mn2e}
\usepackage{graphicx}
\usepackage{amssymb}
\usepackage{bm}

\newcommand{\bq}    {\begin{equation}}
\newcommand{\eq}    {\end{equation}}
\newcommand{\bqr} {\begin{eqnarray}}
\newcommand{\eqr} {\end{eqnarray}}

\title[Growth of magnetic fields in accreting MSPs]{Growth of magnetic fields in accreting millisecond pulsars}

\author[C. Cuofano, A. Drago and G.Pagliara]{C. Cuofano$^{1}$\thanks{E-mail:
cuofano@fe.infn.it (CC); drago@fe.infn.it (AD); pagliara@fe.infn.it (GP)} A. Drago$^{1}$ and G.
Pagliara$^{1}$\\
$^{1}$ Dipartimento di Fisica, Universit\'a di Ferrara 
and INFN sezione di Ferrara, 44100 Ferrara, Italy}

\begin{document}

\date{Accepted --. Received --; in original form --}

\pagerange{\pageref{firstpage}--\pageref{lastpage}} \pubyear{2012}

\maketitle 

\label{firstpage}

\begin{abstract}

R-modes can generate strong magnetic fields in the core of accreting
millisecond neutron stars (NSs).
The diffusion of these fields outside the core causes the growth of the
external magnetic field and thus it affects the evolution of the spin down rates
$\dot{P}$ of the millisecond pulsars (MSPs). 
The diffusion of the internal magnetic field provides a new evolutionary 
path for the MSPs. This scenario could explain the large $\dot{P}$
of the pulsars J1823-3021A and J1824-2452A.

\end{abstract}


\section{Introduction}

In recent papers   
it has been proposed a mechanism based on the r-mode instability
for the evolution of the internal magnetic field 
in millisecond accreting compact stars \citep{Cuofano:2009yg,papernostro,Cuofano2012}. R-modes represent a class of oscillation modes in
rotating neutron stars, which are unstable with respect to the
emission of gravitational waves \citep{Andersson:2000mf}. Only the
existence of efficient damping mechanisms can suppress this
instability and allows neutron stars to rotate at high frequencies. 
An important mechanism for damping is dissipation which can be provided
by shear and bulk viscosities converting the rotational energy of the star
into heat. Another very efficient source of damping is given by
the coupling of the r-modes with the pre-existing poloidal magnetic
field of the star: it has been shown that
r-modes are responsible for the formation of huge internal toroidal magnetic fields
which could reach values of $10^{15}$~G or larger in the case of newly born neutron stars. 
Smaller values are obtained in the case of old and accreting neutron stars.
A large fraction of the rotational energy of the star is thus stored in the magnetic
field~\citep{Rezzolla2000ApJ,Rezzolla:2001di,Rezzolla:2001dh,Sa:2004gn,Sa:2006hn,Cuofano:2009yg}. 
This configuration becomes unstable due to the so-called
Tayler instability \citep{Braithwaite:2009,Cuofano:2009yg}: 
a new poloidal component is generated which can then be 
wound-up itself closing the dynamo loop.
The internal fields evolve into a stable configuration in which the toroidal component
can be significantly stronger than the poloidal \citep{Braithwaite:2009}.

Here we analyze the evolution of the external magnetic field during/after accretion
due to the diffusion outside the core of the magnetic field generated by r-modes.
The increase of the external magnetic field implies a growth of the spin down 
rate $\dot P$ of the millisecond NSs. 
Finally we show that this mechanism can account for the large $\dot{P}$ observed for 
the two pulsars J1823-3021A and J1824-2452A. 

\section{Evolution of the external magnetic field}

The toroidal magnetic fields generated in the core of accreting millisecond neutron stars 
are in the range $B^{tor}_{in}\sim[10^{12}\mbox{--}10^{14}]$~G~\citep{Cuofano:2009yg,Cuofano2012}.
These strong toroidal fields can be stabilized with respect to the
Tayler instability by a much smaller poloidal component ~\citep{Braithwaite:2009} 
that we assume in the range $B_{in}^{pol}\sim[10^{10}\mbox{--}10^{11}]$~G.
We indicate with $t_{0}$ the moment at which the new internal poloidal component is fully developed.

In order for the poloidal field to affect the spin frequency evolution of the 
star it is necessary that the it diffuses outside of the core.  
A crucial issue concerns the time-scale $\tau_{diss}$ on which the internal poloidal 
component can diffuse. 
Unfortunately there are no precise estimates about the value of $\tau_{diss}$ 
in the presence of a layer of superconducting material in the inner crust.
A detailed description of the diffusion of the internal magnetic 
field is beyond the scope of this paper. We limit ourselves to note that 
if $\tau_{diss}$ is shorter than a few ten million years the star enters a propeller regime 
too quickly and it is impossible to accelerate the NS to the higher observed spin frequencies.
On the other hand if $\tau_{diss}$ is much larger than $10^{10}$~yr the expulsion of the magnetic field
would be phenomenologically irrelevant.
It is thus necessary that the diffusion of the
internal poloidal field takes place on a time-scale $\tau_{diss}\approx [10^8\mbox{--}10^{10}]$~yr.
These estimates are compatible with the theoretical predictions given by \citet{Geppert1999} 
and \citet{Konenkov2000MNRAS}.

We begin our analysis at $t=t_{0}$.
We follow the evolution of the external magnetic field 
$B_{ext}^{pol}$ while the internal field $B_{in}^{pol}$ diffuses outside of the core.
For simplicity we assume an exponential growth of the external magnetic field:
\begin{equation}
 B_{ext}^{pol} \,(t) =B_{ext}^{pol} (t_{0}) + B_{in}^{pol} ~[1- \rm exp(-\Delta t/\tau_{diss})]
\label{EvB}
\end{equation}
where $\Delta t = t-t_{0}$ and the initial \textit{external} magnetic field 
is assumed to be $B_{ext}^{pol} (t_0)\sim 10^8$~G. The latter is a rather typical strength for neutron stars 
in Low Mass X-ray Binaries (LMXBs).

\begin{figure}
\begin{centering}
\includegraphics[width=9cm,height=7cm]{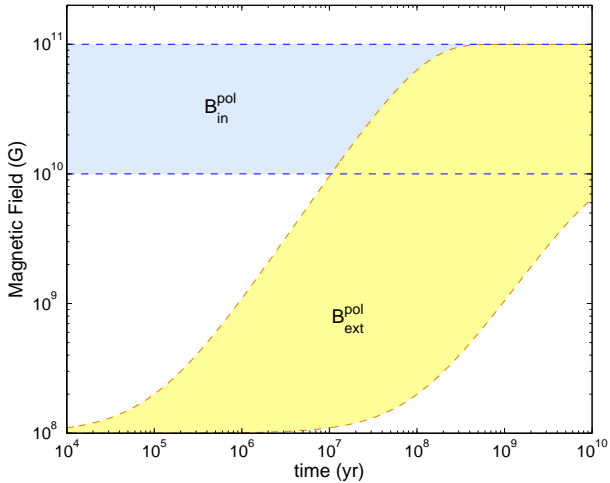}
\caption{(color online) Evolution of the external magnetic field (yellow area) assuming a diffusion time-scale
$\tau_{diss}\sim[10^8\mbox{--}10^{10}]$~yr.
The blue area indicates the value of the internal poloidal component 
$B_{in}^{pol}=[10^{10}\mbox{--}10^{11}]$~G which is about two orders 
of magnitude smaller than the toroidal component $B^{tor}_{in}$. 
Here $B_{ext}^{pol}(t_0)=10^8$~G.}
\label{fig.1}
\end{centering}
\end{figure}

In Figure~\ref{fig.1} we show the evolution of the \textit{external} magnetic field obtained by using Eq.~(\ref{EvB}). 

In the following we consider the diffusion of the new generated internal magnetic field separately 
in the case in which mass accretion is still active and in the case it has terminated.

\subsection{Evolutionary scenario without mass accretion}

Here we consider the evolution of a \textit{recycled} neutron star 
in a scenario in which the mass accretion phase ends 
within a few million years after $t_0$. 
In this case the evolution of the spin period of the star is quite simple because it is not necessary to
take into account the interaction between the accretion disk and the magnetic field during the 
diffusion of $B_{pol}^{in}$.
The total angular momentum of the star satisfies the equation~\citep{Becker2009}:
\begin{equation}
 \frac{dJ}{dt}= - \frac{16}{3}\left(\frac{\pi}{c}\right)^3 \frac{R^6 (B_{ext}^{pol}(t))^2} {P^3}
\label{AngMomV}
\end{equation}
where $c$ is the speed of light, $J\cong I\Omega$ is the angular momentum of the star
and $B_{ext}^{pol}$ is obtained from Eq.~\ref{EvB}. 

In the following we indicate with $P_{0}$ the spin period at the time
$t = t_0$ and we treat $P_{0}$ as a free parameter.
Notice that the internal magnetic field can develop only if
the star enters the r-mode instability region \citep{Cuofano:2009yg}. 
The upper limit of this region is characterized by a period $P_{high}$ 
whose precise value depends on the viscous properties of the star
and on the possible existence of the Ekman layer~\citep{Bondarescu:2007}.
Therefore also the parameter $P_{0}$ is bound by the condition $P_{0}\leq P_{high} \lesssim 5$~ms.

We stress that $B^{pol}_{in}$ and $P_0$ are not the same for every star. 
They actually depend on the very complicated evolution of the internal 
magnetic field which in turn is related to the values of $\dot{M}$ and $B^{pol}_{ext}(t_0)$ 
(see for instance Figure~3 of \citet{Cuofano:2009yg}). 

\begin{figure}
\begin{centering}
\includegraphics[width=9cm,height=7cm]{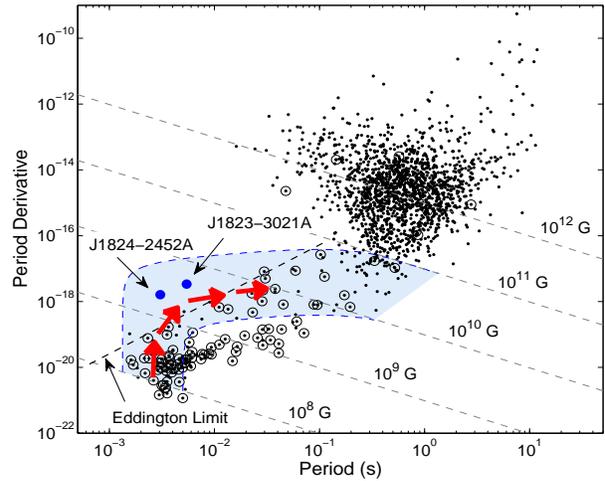}
\caption{(color online) Evolution diagram of accretion-powered X-ray pulsars and of rotation-powered
radio pulsars. The data points are known rotation-powered pulsars; those of pulsars
in binary systems are encircled. The data are from~\citet{dataM}.
The two $anomalous$ stellar objects under study are labeled. 
The dashed line corresponds to stars accreting at the Eddington limit~\citep{Lamb2008}.
The shaded area includes possible paths (indicated by the red arrows for the case without mass accretion) 
of MSPs after $t_0$ (see text).
Notice that the two $anomalous$ stars lie well inside the shaded area.}
\label{fig.2}
\end{centering}
\end{figure}

Results for the evolutionary paths in the $P\mbox{--}\dot{P}$ plane are shown in Figure~\ref{fig.2}.
The shaded strip includes the possible trajectories of the stars 
(the red arrows are included as guidelines for the temporal evolution). 
The shape of this strip is regulated mainly by the three parameters 
$\tau_{diss}$, $B^{pol}_{in}$ and $P_{high}$ ($B^{pol}_{ext}(t_0)$ plays a marginal role). 
The width of the strip at the beginning is fixed by $P_{high}$ and $P_{low}$.
The latter parameter corresponds to the minimal period reachable by a recycled MSP.
Observationally its value is of the order of 
$P_{low}\approx 1.3$~ms~\citep{Chakrabarty2003Natur,Chakrabarty2005ASPC}\footnote{This value 
can be justified theoretically by considering
the deformation of the star generated by the internal toroidal field \citep{Cuofano2012}.}.
$B^{pol}_{in}$ fixes the width 
of the strip at late times and $\tau_{diss}$ regulates the height of the elbow: 
the smaller $\tau_{diss}$ the higher the elbow. 

Remarkably, we can populate the region above the line corresponding 
to the Eddington limit (dashed line). This is exactly the region 
where J1823-3021A and J1824-2452A lie. 

\subsection{Evolutionary scenario with mass accretion}
\label{accretion}

If the mass accretion phase lasts tens or hundreds of million years
after $t_0$, the evolutive scenario is more complicated.

The plasma flowing onto the neutron star forms an accretion disk whose
inner edge is given by the coupling radius of the magnetosphere 
(of the same order of magnitude of the Alfv\'en radius)
$r_{\rm mag}\sim B^{4/7}R^{12/7}(\dot{M}\sqrt{2GM})^{-2/7}$~\citep{Tauris2012}.
The interaction between the magnetic field of the compact object and
the conducting material flowing from its companion can provide the necessary torque
$\dot{J}_a=\dot{M}(GMr_{\rm mag})^{1/2}$ to spin up the pulsar~\citep{shapiro}.
However, under certain conditions, it is possible that this interaction leads to a torque reversal. 
The growth of the external field $B_{ext}^{pol}$ causes the magnetic boundary $r_{\rm mag}$
to move outward relative to the corotation radius $r_{\rm co}=(GM/\Omega^2)^{1/3}$ 
defined as the distance at which the spin frequency 
of the star is equal to the Keplerian frequency.
The star enters a propeller phase when $r_{\rm mag}>r_{\rm co}$: a centrifugal barrier
prevents the material to flow onto the star and a new braking 
torque $\dot{J}_M\sim - \dot{M}(GMr_{\rm mag})^{1/2}$ acts to slow down the pulsar~\citep{Tauris2012}.
\begin{figure}
\begin{centering}
\includegraphics[width=9cm,height=7cm]{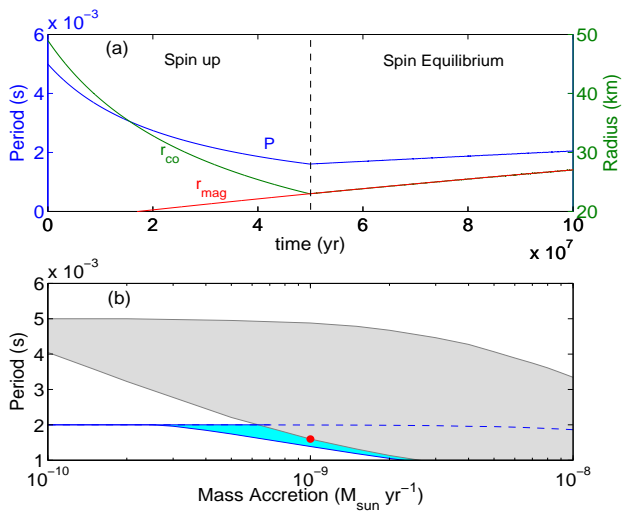}
\caption{(color online) \textit{Panel (a)}: time evolution of $r_{co}$ (green line), 
$r_{mag}$ (red line) and period $P$ of the star (blue line).
The dashed black line indicates the moment at which the spin equilibrium phase takes place.
Here we assume $P_{0}=5$~ms, $B_{\rm in}^{\rm pol}=10^{10}$~G and
$\dot{M}=10^{-9}\,\rm M_\odot\rm yr^{-1}$.
\textit{Panel (b)}: frequencies at which the spin equilibrium phase 
takes place as a function of the mass accretion rate. 
We show two examples, corresponding to $P_0= 5$~ms (gray area) and $P_0 = 2$~ms (cyan area). 
The width of the strips corresponds to the uncertainty on $B_{pol}^{ext}$ 
(see yellow area in Figure~\ref{fig.1}). The red dot refers to the case
shown in panel (a). 
}
\label{fig.3}
\end{centering}
\end{figure}

In Figure~\ref{fig.3}a we show a typical evolutionary path for an
accreting neutron star if the diffusion of the internal poloidal component $B_{in}^{pol}$
is taken into account. 
The growth of the external magnetic field $B_{\rm ext}^{\rm pol}$ leads to the
end of the spin-up phase.
The star evolves into an equilibrium spin phase, it alternates 
between spin-up and spin-down ($r_{\rm mag}\sim r_{\rm co}$) 
and finally it appears as a radio pulsar at the end of the mass accretion phase.

Finally, from the analysis presented above we can notice that in order to reach high
spin frequencies it is necessary a mass accretion rate
$\dot{M}\gtrsim 10^{-9}\, \rm M_\odot \rm yr^{-1}$ (see Figure~\ref{fig.3}b).

\section{Millisecond Pulsars J1823-3021A and J1824-2452A}

In the previous Section we have presented a general scenario for the
evolution of MSPs when the production and the diffusion
of the internal magnetic field is taken into account. In this Section we concentrate on the cases of the two isolated
pulsars J1823-3021A and J1824-2452A. 

In the most widely discussed model, MSPs have been re-accelerated through 
mass accretion, in particular by having spent a fraction of their lives as part of a LMXB. 
The way by which they end up as isolated pulsars is still not fully understood.  
A possibility is that the companions are destroyed through ablation caused by energetic 
radiation from the pulsar \citep{Bhattacharya1991}. Another possibility requires mass loss 
from the companion which can take place in very close binary systems due to 
tidal dissipation \citep{Radhakrishnan1986Ap}.
A totally different scheme to produce an isolated MSP is the one in which the 
pulsar had an interaction with another star or binary in the globular cluster, 
and was spun-up in the process \citep{Lyne1996ApJ}.
In the following we will concentrate on the scenario in which MSPs are
produced in LMXBs.
\\

The pulsar J1823-3021A, located in the globular cluster NGC 6624, was
discovered in 1994~\citep{biggs1, biggs}. The Fermi Large
Area Telescope has recently detected the $\gamma$ ray counterpart of
its pulsations~\citep{science} with a luminosity of $L_{\gamma}=8.4 \pm
1.6 \times 10^{34}$ ergs s$^{-1}$, the highest observed $\gamma$ ray
luminosity for any MSP. The pulsation period $P$ is $5.44$ ms and the
observed spinning down rate is extremely large $\dot{P}_{obs}=+3.38
\times 10^{-18}$ s s$^{-1}$, larger by far than the typical value for
other MSPs. These properties make J1823-3021A an extremely interesting stellar object. 
The total observed $\gamma$ ray emission implies that a significant fraction of $\dot{P}_{obs}$ is
due to the intrinsic spinning down rate $\dot{P}$~\citep{science}. 
Within the standard magnetic dipole model for pulsar emission, such a large $\dot{P}$ is
generated by a large surface magnetic field $B_0$, larger than
$10^9$ G. This fact represents a puzzle
for standard MSP theory~\citep{science}: if J1823-3021A was spun-up
by mass accretion from a companion, due to its large magnetic field
$B_0$, even assuming an accretion rate at the Eddington limit $\dot
M_{Edd}=10^{-8} M_{\odot}/yr$, the star would have reached a
period $P=6.6$ ms, which is larger than its present period.
\\

The isolated millisecond pulsar J1824-2452A is located in the globular cluster 
M28 at $5.6$~kpc from the Sun \citep{lyne}. The pulsation period is 
$P=3.054$~ms and the observed spinning down rate is $\dot{P}_{obs}=+1.62
\times 10^{-18}$ s s$^{-1}$. Indirect evidence exists that this large $\dot{P}$ observed
is indicative of the true value, e.g. the detection of glitches \citep{Cognard2004ApjL},
giant pulses \citep{Knight2006ApJ}, strong X-ray pulses \citep{Rots1998ApJ} and a large
second period derivative $\ddot{P}$ \citep{Cognard1996ApJ}. As for J1823-3021A, the large
$\dot{P}$ implies a surface magnetic field $B_0> 10^9$~G and the combination
of values ($P\mbox{--}\dot{P}$) is difficult to explain in the standard MSPs theory.
\\

Note that these anomalous pulsars can be explained
in the standard model if the accretion occurred at super-Eddington rate.
About such a possibility, we note that all the accreting compact stars detected 
up to now in LMXBs
have steady mass accretion rates $\dot{M}\ll\dot{M}_{Edd}$~\citep{Galloway:2007bt}\footnote{LMXBs 
are often transient sources and may have short-term active periods of intense accretion;
large $\dot{M}$ presented in \citet{Galloway:2007bt} in parentheses refer to these active
periods and not to steady accretion.}.
Moreover, taking into account also the interaction between the external magnetic 
field and the accretion disk, a minimal value for the mass accretion rate
$\dot{M}\gtrsim 5\times 10^{-8}~\rm M_{\odot}\,\rm yr^{-1}$ 
is necessary to accelerate the stars up to the present spin frequency. 
For lower values of $\dot{M}$ the accreting NS enters a propeller regime
that prevents a further acceleration.

In our analysis we have presented a scheme in which the MSPs can populate
the region occupied by J1823-3021A and J1824-2452A. It is important to remark that
the existence of a propeller phase does not invalidate the possibility to interpret
these two special MSPs in our scheme. 
Indeed, in Figure~\ref{fig.3}a we show that the spin frequency of the star
does not change significantly during the spin equilibrium phase:
even if a fast spin-down of the star takes place during the propeller regime 
\citep{Tauris2012}, the star nevertheless will continue to appear as a radio pulsar 
in the shaded area of Figure~\ref{fig.2}. Concerning the two anomalous stellar 
objects they are no more accreting and 
if a propeller phase took place during accretion it is clearly already over.
The exact position in the $P$-$\dot P$ plane before a possible propeller phase 
cannot be determined precisely, but it was located at higher frequencies: the existence 
of this phase would make 
therefore even more difficult to interpret these two objects within the standard pulsar model
(because they would lie further from the Eddington limit) but it is easy to accommodate in our model.

\section{Conclusions}

We have discussed the evolution of the external magnetic field
of recycled MSPs due to the diffusion of strong internal magnetic
fields generated by r-modes during the mass accretion phase.
Our analysis is based on the previous work of \citet{Cuofano:2009yg} 
and \citet{papernostro} where it has been shown that 
r-modes generate strong internal magnetic fields in accreting
compact stars. 
We have shown that the growth of the external magnetic field
affects the evolution of the spin-down rate $\dot{P}$ of the MSPs and that this mechanism
can explain the high $\dot{P}$ of J1823-3021A and of J1824-2452A.
Our analysis open new evolutionary paths for the \textit{recycled} MSPs
in the $P\mbox{--}\dot{P}$ plane.

\section*{acknowledgements}
G.P. acknowledges financial support from the Italian Ministry of
Research through the program Rita Levi Montalcini.

\label{lastpage}

\end{document}